\documentclass[%
 reprint,
 amsmath,amssymb,
 aps,
]{revtex4-2}
\usepackage{hyperref}                         
\usepackage[all]{hypcap}                      
\usepackage{graphicx}
\usepackage{dcolumn}
\usepackage{bm}
\usepackage{xcolor}
\usepackage{caption}
\usepackage{subcaption}
\usepackage[font=small,labelfont=bf,justification=justified]{caption}

\usepackage{float}
\usepackage[english]{babel}

\begin{document}

\preprint{APS/123-QED}

\title{Density probabilities and quantum critical phenomena of a Bose-Fermi Mixture\\ in 1D Double well potential}

\noindent
\author{R. Avella $^1$}
\altaffiliation[]{rgavellas@unal.edu.co}
\author{J. Nisperuza$^1$, JP Rubio$^1$ and D. Grajales$^2$}\\
\affiliation{%
$^1$Fundaci\'{o}n Universitaria los Libertadores\\ Faculty of Engineering and Basic Sciences\\ Department of Aeronautical Engineering\\ A. A. 75087 Bogot\'{a}, Colombia.\\
}
\affiliation{%
$^2$ Universidad EAN\\ Faculty of Engineering \\ A. A. 110221 Bogot\'{a}, Colombia.\\
}
%
%
%
%
%
%
%

\date{\today}

\begin{abstract}
The time evolution of probability density, the ground-state fidelity and the entanglement of a Bose-Fermi mixture in a 1D double well potential, are studied through the two mode approximation.
We found that the behaviour of the quantum return probability shows three distinct regions. The first region is characterized by a complete miscibility, and correlated tunneling of bosons and fermion. The second region is characterized by correlated sequential tunneling and in the last region we found an increase in the tunneling frequency of the two species.

We found through the Von Neumann entropy, that the boson-fermion coupling allows a maximum entanglement of quantum correlations of bosons and fermions in the same value. Finally we calculated the fidelity in the $\lambda_{FF}-\lambda_{BF}$ and $\lambda_{BB}-\lambda_{BF}$  planes and we found that the drop of the two fidelities becomes deeper and deeper as the boson-fermion interaction decreases.
\end{abstract}

\maketitle


\section{INTRODUCTION}
The perfect control over the trapping geometries in ultracold gases experiments allow to confine different numbers of atoms in different lattice sites through magnetic Feshbach resonance or confinement induced resonance \cite{Ospelkaus06, Hadzibabic03, Inouye04}, both the effective intra and inter component interactions, can be tuned with great precision.
The perfectly controllable physical parameters\cite{Anglin02}, provides without precedents platforms in experimental study quantum many-body physics, for instance,  interesting quasi-one-dimension experiments have been realized in harmonic \cite{Pitaevskii03, pethick_smith_2001}, double-well\cite{PhysRevLett.95.010402, Gati_2007}, periodic\cite{PhysRevLett.74.1542}, and bichromatic\cite{PhysRevA.80.023606} optical-lattice traps.

Dimers and single double-well system has been realized in one, two and three dimensions \cite{Trotzky08, PhysRevLett.98.200405, PhysRevA.76.043606} and allow to study the notion of state as a linear superposition of ‘classical’ states\cite{Feynman66}, where the system can reside in a superposition of two or more degenerate states\cite{Holstein88, Chebotarev98, Garg00}. In these systems the tunneling between the local double-well potentials is negligible compared to tunneling inside the double well potential and each site have a well-defined and almost identical quantum state \cite{Anderlin07, Trotzky08,PhysRevA.73.033605};  this phenomenon has applications in solid-state devices, solar cells and microscopes \cite{wiesendanger_1994}.

In quantum information processing, these systems have been used as a way to make quantum logic gates \cite{PhysRevLett.98.070501}, as well as, the bosonic Josephson junction \cite{PhysRevA.55.4318, PhysRevLett.79.4950, PhysRevLett.95.010402}, squeezing and entanglement of matter waves \cite{PhysRevLett.98.030407,Esteve08}, matter wave interference \cite{ANDREWSC97, Schumm06}, exact many-body quantum dynamic in one dimension and Josephson effect \cite{PhysRevA.79.033627, PhysRevLett.93.120401, refId0, PhysRevA.78.041403, PhysRevA.61.031601, Cataliotti01, PhysRevLett.95.010402}.

The influence of fermions onto bosons has been investigated in different mixture, for example in $^4He-^3He$ \cite{Pollet08, McNamara06}, $^{87}Rb-^{40}K$ \cite{Klempt08, Karpiuk06},  $^{41}K-^{6}Li$\cite{Lous18},  $^{87}Rb-^{40}K$ \cite{Wille06},  $^{170}Yb-^{173}Yb$,  $^{174}Yb-^{173}Yb$ \cite{Sugawa11}. Other Bose-Fermi mixtures  (BFM) that have been experimentally reported are $^7Li-^6Li$ \cite{Akdeniz02}, $^{39}K-^{40}K$ and $^{41}K -^{40}K$ \cite{Vichi98}.

The Bose-Fermi mixture (BFM) have been studied numerically \cite{Roth02, Liu03, Modugno03, Adhikari04}, semi-analytically \cite{Miyakawa01, Karpiuk05}, or in the Thomas–Fermi approximation\cite{Pelster07}. The Bose-Fermi interaction induces the pairing of fermions \cite{Bijlsma00, Heiselberg00, Viverit02},  boson phase transition from the Mott insulator to super fluid \cite{Mering08, Bukov14, Fehrmann04} and the spontaneous symmetry breaking of a superfluid \cite{Adhikar10}. Other studies indicate an asymmetry between the attraction and repulsion cases \cite{Best09, Albus03}, as well as phase separation, spatial modulation \cite{Polak10}, supersolid phase and charge density wave \cite{Titvinidze08}.

The critical phenomena in quantum many-body systems is studied by means of quantum information theory concepts, for example the entanglement and quantum fidelity. The first has been related to quantum phase transitions \cite{PhysRevA.66.032110, PhysRevA.70.052302}, the latter as a measure of similarity between ground states \cite{PhysRevLett.96.140604, PhysRevB.75.014439, PhysRevLett.98.110601}.

In this paper we study the time evolution of probability density, the ground-state fidelity and the entanglement of a Bose-Fermi mixture in a 1D double well potential considering spinless bosons in the soft-core limit and spin one half fermions. We consider the boson-boson, fermion-fermion and boson-fermion interaction to be of the repulsive type at zero temperature and considering that both species have the same mass. This system will be carried out by means of the two-mode model in a double well potentia, using the lowest symmetric and antisymmetric wave functions. This model produces the best agreement with experimental results and numerical solutions of the time-dependent Gross-Pitaevskii equation in 1D and 3D \cite{Ananikian05,Ostrovskaya00,Rey03}.

The paper is organized as follows. The model used to describe a mixture of bosonic and fermionic atoms is introduced in Sec.\ref{sec:Bose-Fermi mixtures model}. In section
\ref{sec:probabilities}, we vary the inter and intra species interactions to study the temporal evolution of probability densities and we make a brief introduction to quantum critical phenomena that occur in the system under study in Sec. \ref{sec:QuaCriPhe}.  Finally in Sec. \ref{sec:Conclusions} we make remarks.

\section{Physical Model}\label{sec:Bose-Fermi mixtures model}
\subsection{Double well potential}
Due to the experimental possibility of confining quantum gases in an array of many copies of the double well system,  we considered the study in one of these potentials, using the experimental setup of \cite{PhysRevLett.92.050405}, where the radial separation of the potential wells is $d=13\mu m$, a trap depth of $h\times4.7Khz$ and width of each well of $a=6\mu m$. This potential confines isotopes of $^{170}Yb$ (bosons) and one half spin isotopes of $^{171}Yb$ (fermions), allowing a density of fermions per site between $0\leq\rho_F\leq2$ and a maximum number of scalar bosons of $n_{max}^B=2$. This value is due to the fact that in several reports has been found that the qualitative physical properties obtained for $n_{max}^B=2$ do not change when $n_{max}^B$ is increased \cite{Pai96, Rossini12}.

The two-particle Hamiltonian for the system in one dimension is
\begin{equation}
\label{hubofear}
\begin{split}
\hat{H}_{BF}=\hat{H}_{B}^{\lambda_{B}}(x_1,x_2)+\hat{H}_{F}^{\lambda_{F}}(x_1,x_2)+\lambda_{BF}\delta(x_1-x_2),
\end{split}
\end{equation}
where
\begin{equation}
\begin{split}
&\hat{H}_{B}^{\lambda_{B}}(x_1,x_2)=\\
&-\frac{\hbar^{2}}{2m_{B}}\Big(\frac{d^{2}}{dx_1^{2}}+\frac{d^{2}}{dx_2^{2}}\Big)+V(x_1)+V(x_2)
+\lambda_{B}\delta(x_1-x_2),
\end{split}
\end{equation}
and
\begin{equation}
\begin{split}
&\hat{H}_{F}^{\lambda_{F}}(x_1,x_2)=\\
&-\frac{\hbar^{2}}{2m_F}\Big(\frac{d^{2}}{dx_1^{2}}+\frac{d^{2}}{dx_2^{2}}\Big)+V(x_1)+V(x_2)
+\lambda_{F}\delta(x_1-x_2).
\end{split}
\end{equation}
The mass of $^{170}Yb$ ($^{171}Yb$) is represented by means of $m_B(_F)$, $V(x)$ is the external confinement potential that is the same for both species. The repulsive contact potential between bosons (fermions) is represented by delta-function potential $\lambda_{B}(_F)\delta(x_1-x_2)$, where $\lambda_{B(F)}>0$ and $\lambda_{BF}\delta(x_1-x_2)$ is the repulsive interaction between two ultracold neutral atoms of different statistic.

We consider that the perturbation is sufficiently small, therefore is considered that the amplitudes of other states do not mix \cite{doi:10.1119/1.3583478}, so we use the projection of the wave function onto two states that is used in the study of BEC in a double-well potential and in the investigation of Fermi super fluid \cite{PhysRevA.77.043609}.
The different contributions of these terms can be obtained in a general way from
\begin{equation}
\begin{split}
\label{int}
&\lambda\langle\Psi_{i,j}|x_1,x_2\rangle\delta(x_{1}-x_{2})\langle x_1, x_2|\Psi_{k,l}\rangle=\\
&\lambda\iint\Psi_{i,j}^*(x_{1},x_{2})\Psi_{k,l}(x_{1},x_{2})\delta(x_{1}-x_{2})dx_{1}dx_{2},
\end{split}
\end{equation}
where $\Psi_{i,j}(x_{1},x_{2})$ is the two particle wave function and $\Psi_{i,j}^*(x_{1},x_{2})$ is its complex conjugate.

Since the function $V(x)$ is even, a basis of eigenvectors of $\hat{H}_{B(F)}$can be found wich are even or odd. The wave functions of these vectors are symmetrical (s) and antisymmetrical (a) linear combinations

\begin{equation}
\label{symmetrical}
\psi^{n}_{s(a)}(x)=\frac{\psi_1^n(x)\pm\psi_2^n(x)}{\sqrt{2}},
\end{equation}

Where $\psi_{1,(2)}^n(x)$ is a one particle states that is twofold degenerate and superscript n indicate the nth energy value.  In the states $\psi^{n}_{s}(x)$ and $\psi^{n}_{a}(x)$, the particle can be found in the rigth $|\psi_{R}(x)\rangle$ or in the left $|\psi_{L}(x)\rangle$ of the double potential well respectively.


Finally we adopt our units of length, $l=1\mu$, energy $E_a=E/\xi$ with $\xi=10^{-31}$ and time $\tau=\hbar/\xi$\cite{Avella_2016}. Henceforth, we will measure lengths, energies and time in these
units.

\subsection{Bose and Fermi mixture in a double well}
The previous sections have prepared all the tools we need to understand the behavior of our particles in a double well.  The requirements of overall exchange antisymmetry for fermions and symmetry for bosons, introduce a connection between the spin that has influence on the occupancy of the energy levels and spatial wave functions, due to the particles are identical. In this way the subset of all the possible two-particle wave functions that has overall antisymmetry with respect to the exchange of particle labels as required for identical fermions are linear combinations of the terms
\begin{equation}
\Psi_s^{space}\Psi_a^{spin} \:\:  and \:\:  \Psi_a^{space}\Psi_s^{spin}.
\end{equation}
There is a singlet for $\Psi_a^{spin}$ and a triplet for $\Psi_s^{spin}$ making a total of four functions.

\noindent
Singlet
\begin{equation}
\label{FerFunSin}
\frac{|\psi_{L}(x_1)\psi_{R}(x_2)\rangle-|\psi_{R}(x_1)\psi_{L}(x_2)\rangle}{\sqrt{2}}\:\:\frac{|\uparrow\downarrow\rangle-|\downarrow\uparrow\rangle}{\sqrt{2}}
 \end{equation}

\noindent
 Triplet
\begin{equation}
\label{FerFunTri}
\begin{cases}
|\psi_{L}(x_1)\psi_{L}(x_2)\rangle&|\uparrow\uparrow\rangle
 \:\:\:\:\\\
\frac{|\psi_{L}(x_1)\psi_{R}(x_2)\rangle+|\psi_{R}(x_1)\psi_{L}(x_2)\rangle}{\sqrt{2}}&\frac{|\uparrow\downarrow\rangle+|\downarrow\uparrow\rangle}{\sqrt{2}}
 \:\:\:\:\\\
|\psi_{R}(x_1)\psi_{R}(x_2)\rangle&|\downarrow\downarrow\rangle
\end{cases}
 \end{equation}

The wave function is symmetric with respect to exchange of the particle labels for two bosons, so that the wave functions are linear combinations
of the terms,
\begin{equation}
\Psi_s^{space}\Psi_s^{spin} \:\:  and \:\:  \Psi_a^{space}\Psi_a^{spin}.
\end{equation}
For the case under study, bosons with spin 0 are considered, so that there is a total of three functions.

\begin{equation}
\label{BosFunTri}
\begin{cases}
|\psi_{L}(x_1)\psi_{L}(x_2)\rangle \:\:\:\:\\\
\
\frac{|\psi_{L}(x_1)\psi_{R}(x_2)\rangle+|\psi_{R}(x_1)\psi_{L}(x_2)\rangle}{\sqrt{2}}\:\:\:\:\\\
\
|\psi_{R}(x_1)\psi_{R}(x_2)\rangle
\end{cases}
 \end{equation}

\section{Bose-Fermi probabilities}\label{sec:probabilities}
The wave function of the two particles at time $t$ is obtained by assuming that at $t = 0$, the two particles are in right ($\Psi_{RR}(x_1,x_2,t)$) side of double well potential (symmetric state of equation  \ref{symmetrical})

\begin{equation}
\label{twowave}
\begin{split}
&\Psi_{RR}(x_1,x_2,t)=\\
&\frac{e^{-i\frac{E_s^1+E_a^1}{\hbar}}}{2}\Big[e^{i\Omega_1t}\psi_s^1(x_1)\psi_s^1(x_2)+ \psi_s^1(x_1)\psi_a^1(x_2)\\
&+\psi_a^1(x_1)\psi_s^1(x_2)+e^{i\Omega_1t}\psi_a^1(x_1)\psi_a^1(x_2)\Big],
\end{split}
\end{equation}
where $\Omega_1=\frac{E_a^1-E_s^1}{\hbar}$ is the Bohr frequency.

From the two particles wave function \ref{twowave} we get the probability density for the system $|\Psi_{RR}(x_1,x_2,t)|^{2}$.

The initial state of our system, is configured with two spinless soft core bosons and two one half spin fermions in the right side of the double well potential.
This configuration is mainly due to the fact that we want to explore entanglement and fidelity  when varying  the boson-boson $\lambda_{BB}$, fermion-fermion $\lambda_{FF}$ and boson-fermion $\lambda_{BF}$ repulsive contact interaction terms. Since we are working with a two mode model, the inter and intra particle interaction, must be small enough to prevent the amplitudes of the two lower energy modes mixing with other states.

\begin{figure}[H]
\centering
\begin{subfigure}[b]{1\linewidth}
\includegraphics[width=\linewidth]{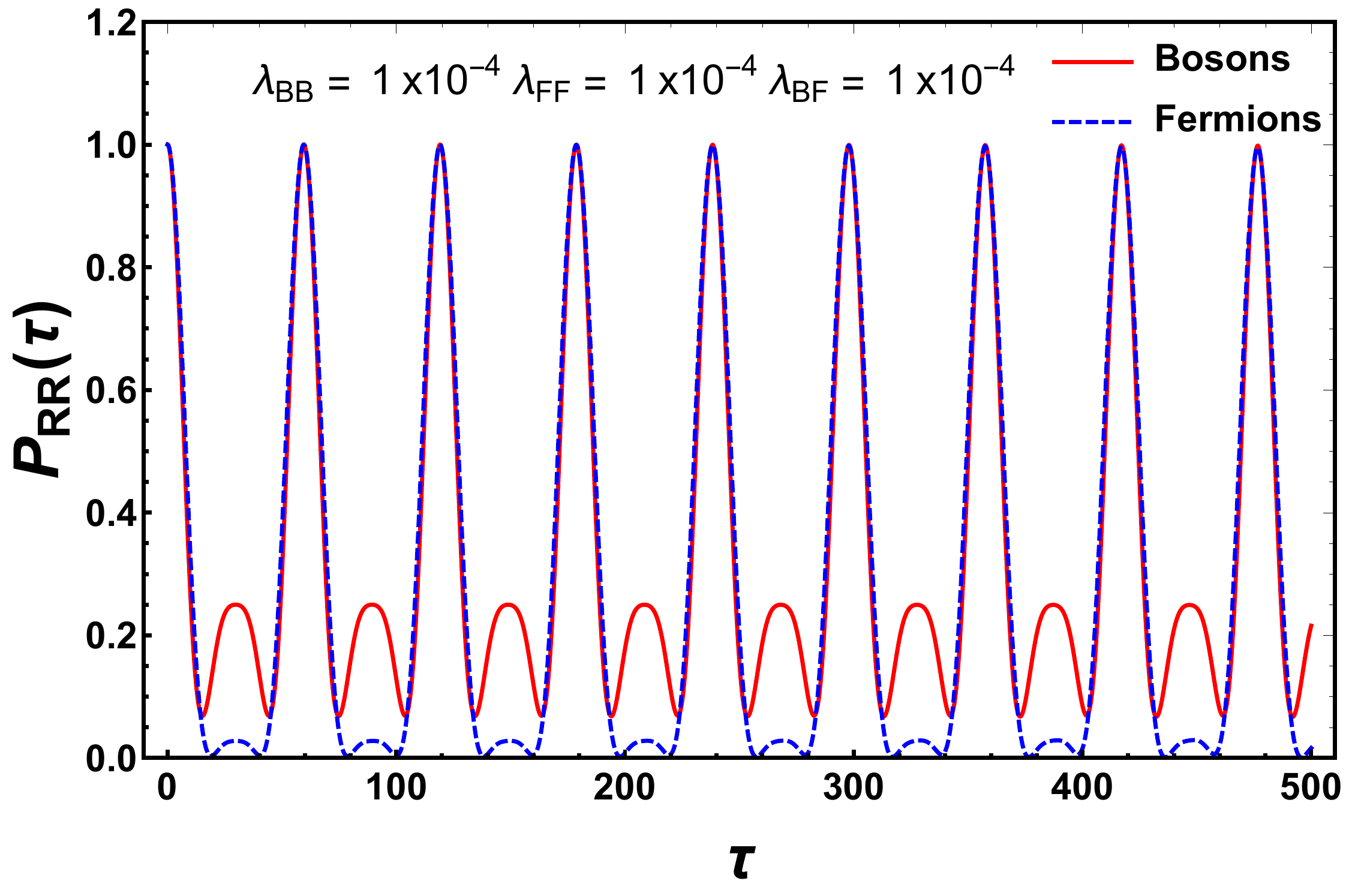}
\captionsetup{justification=justified}
\caption{Periodic density probabilities by bosons (red line) and fermions (blue dashed line).}
\label{symmetry1}
\end{subfigure}

\begin{subfigure}[b]{1\linewidth}
\includegraphics[width=\linewidth]{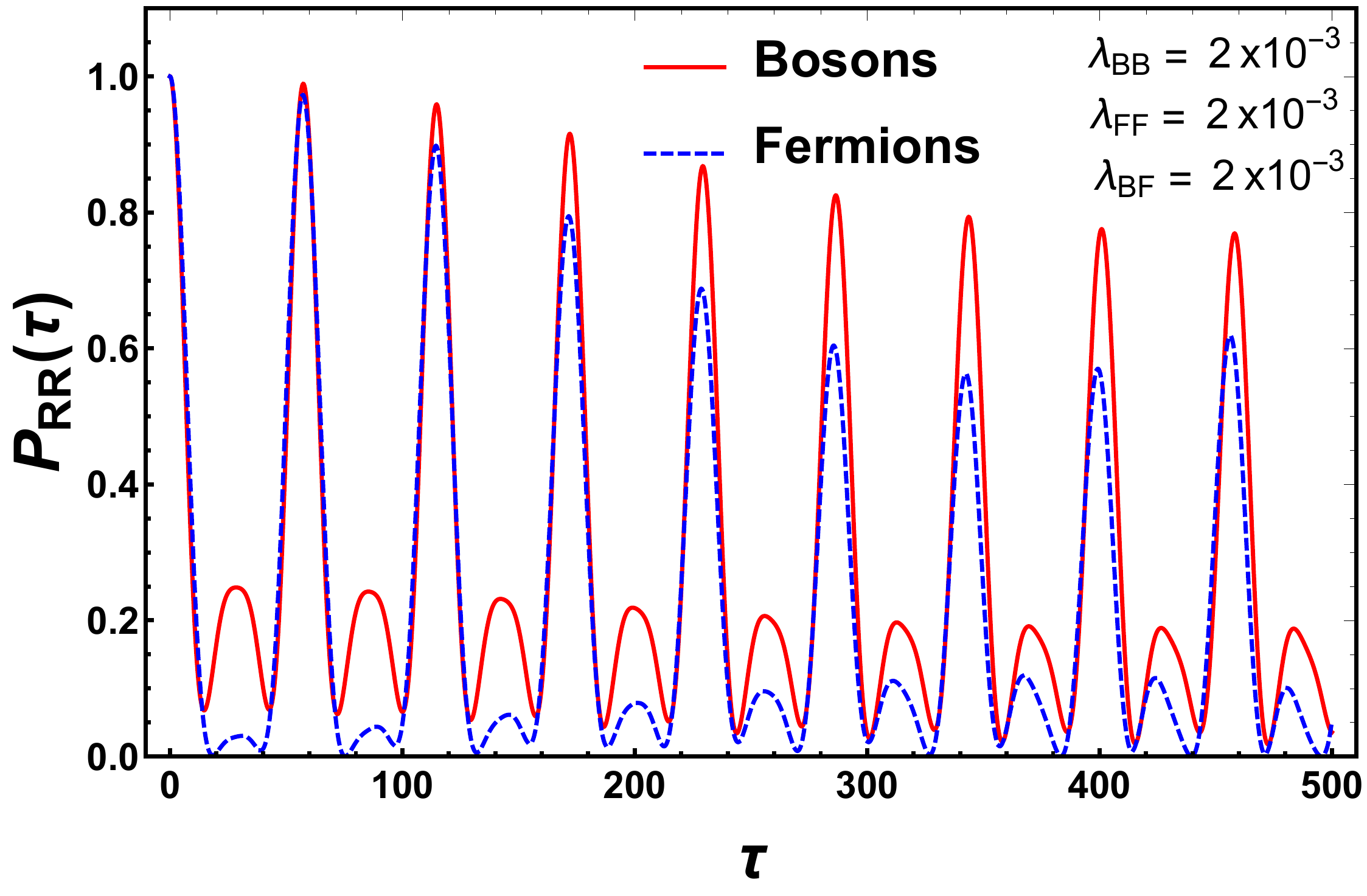}
\captionsetup{justification=justified}
\caption{Slowly dampens for the initial conditions of  the probability density of both bosons (red line) and fermions (blue dashed line) as a function of time.}
\label{symmetry2}
\end{subfigure}

\caption{Time evolution of bosons and fermions density probabilities on the right side of the double well $P_{RR}(\tau)$, as a function of the dimensionless parameter of time $\tau$.}
\label{symmetry}
\end{figure}

The behaviour of the quantum return probability $P_{RR}$ presents three distinct regions. The first region is characterized by correlated tunneling of bosons \cite{Dutta15} and fermions\cite{PhysRevB.72.035330, YangYong22}, where probabilities are periodic and correspond to very small values of the effective coupling constants, with orders of magnitude between $10^{-5}$ and $10^{-4}$  as illustrated in figure \ref{symmetry1}, where the bosonic (red line) and fermionic (blue dashed line) probability densities exhibit complete miscibility  \cite{jiang19, PhysRevA.97.053626} on the right-hand side of the double well, characterized by an overlapping of the probability densities of the two species as a function of time. It can also be seen that the bosons do not perform complete tunneling and the fermions exhibit a small damping.

This periodical behavior is maintained until $U_{BB}=U_{FF}=U_{BF}=4\times10^{-4}$, from where a slowly damping of the probability densities starts to be generated, which increases as the interaction parameters increases  as illustrated in figure \ref{symmetry2}. The figure shows that due to the increasing in the interaction parameters, the probability density of both bosons and fermions decreases as a function of time, indicating that the particles tend to separate and the tunneling frequency increases characterized by a loss of miscibility as a function of time, as well as a tendency for bosons to tunnel completely and a loss of damping in fermionic tunneling.

\begin{figure}[H]
\centering
\begin{subfigure}[b]{1\linewidth}
\includegraphics[width=\linewidth]{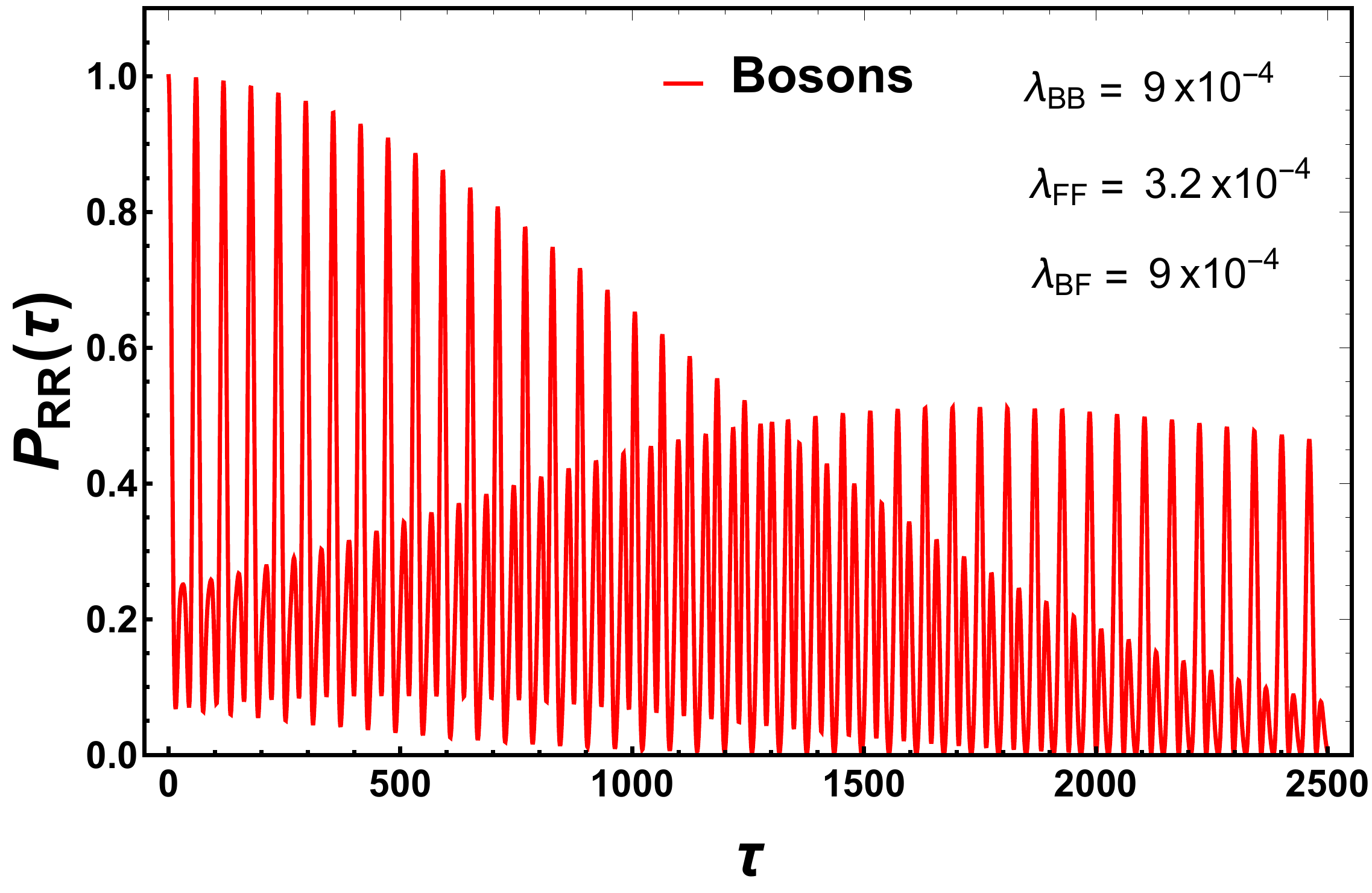}
\caption{Bosons correlated sequential tunneling through a double barrier, as a function of the dimensionless parameter of time $\tau$.}
\label{bosonscte}
\end{subfigure}

\begin{subfigure}[b]{1\linewidth}
\includegraphics[width=\linewidth]{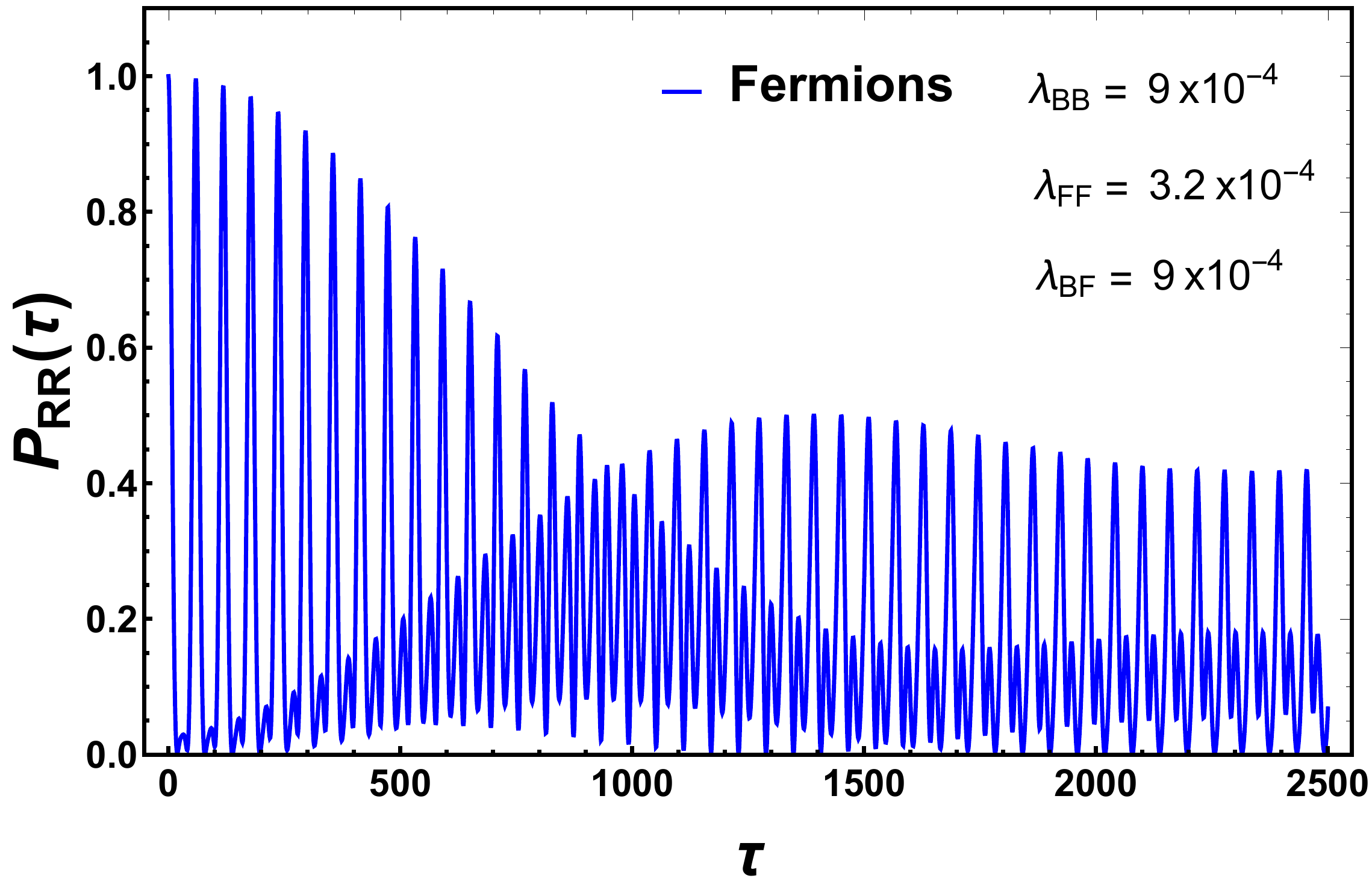}
\caption{Fermions correlated sequential tunneling through a double barrier, as a function of the dimensionless parameter of time $\tau$.}
\label{fermionscte}
\end{subfigure}

\caption{Slowly dampens for the initial conditions of  the probability density of both bosons (red line) and fermions (blue line) as a function of time for: $\lambda_{BB}=9\times10^{-4}, \lambda_{FF}=3.2\times10^{-4}$ and $\lambda_{BF}=9\times10^{-4}$.}
\label{constant2}
\end{figure}

The second region found indicates that due to the increasing in the interaction parameters, the particles of both species tend to separate and occupy both sides of the confinement potential and give rise to correlated sequential tunneling through a double barrier, characterized by having one particle of each species on each side of the well and which remains constant at dimensionless time interval  of  $1275.9\leq\tau\leq2283.19$ for bosons  and $1156.32\leq \tau\leq1688.38$ for fermions, as shown in the figure\ref{constant2}. It was also found that the fermionic probability density tends to damp more rapidly than the bosonic one, as illustrated in figures \ref{bosonscte} and \ref{fermionscte} respectively.

\begin{figure}[H]
\centering
\begin{subfigure}[b]{1\linewidth}
\includegraphics[width=\linewidth]{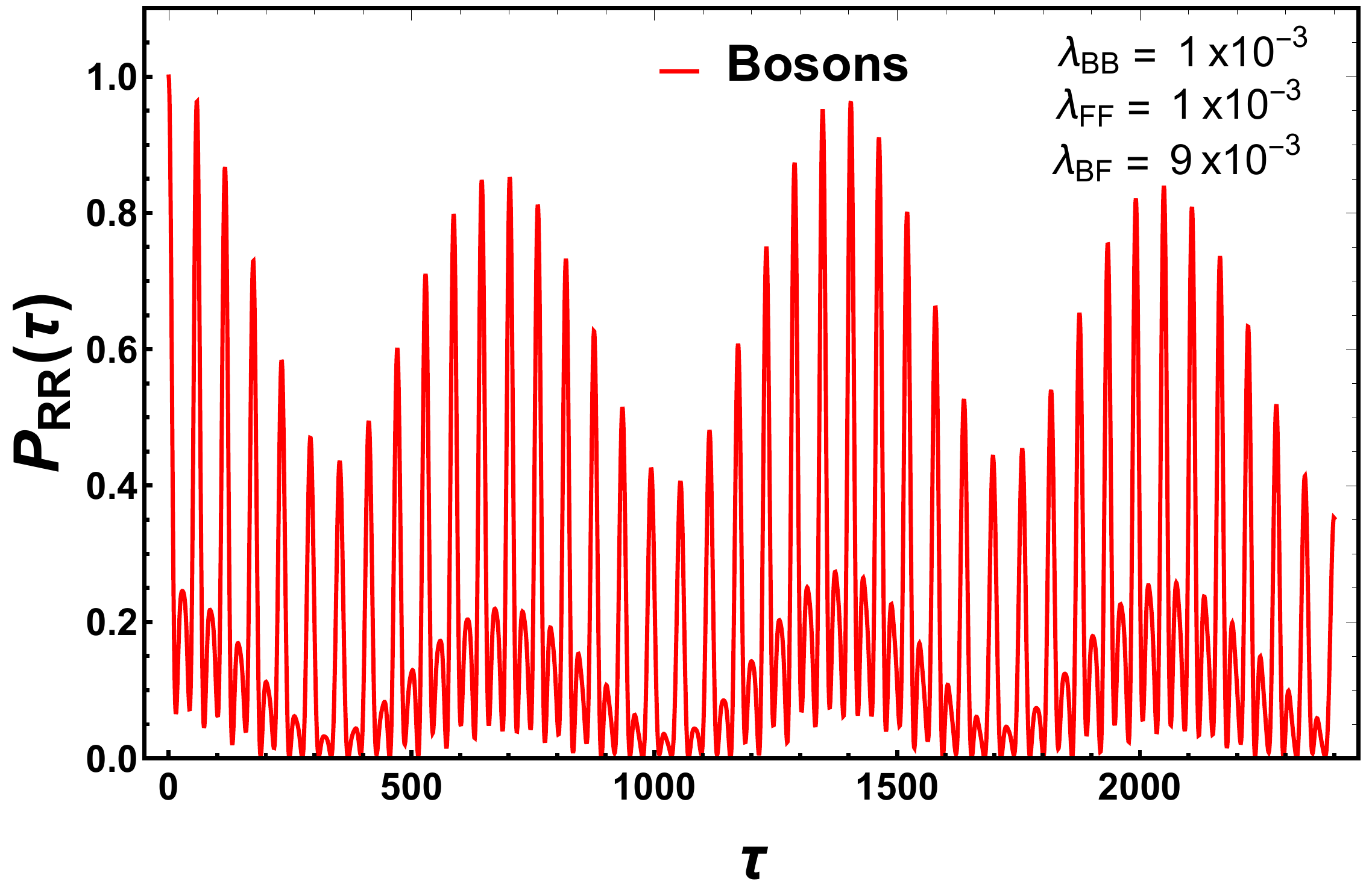}
\caption{Quasi periodicity for bosons. The graph shows $\tau$ for which the maximum occupancy per site is two bosons and $\tau$ for which sequential tunneling occurs.}
\label{bosonscuasi}
\end{subfigure}
\begin{subfigure}[b]{1\linewidth}
\includegraphics[width=\linewidth]{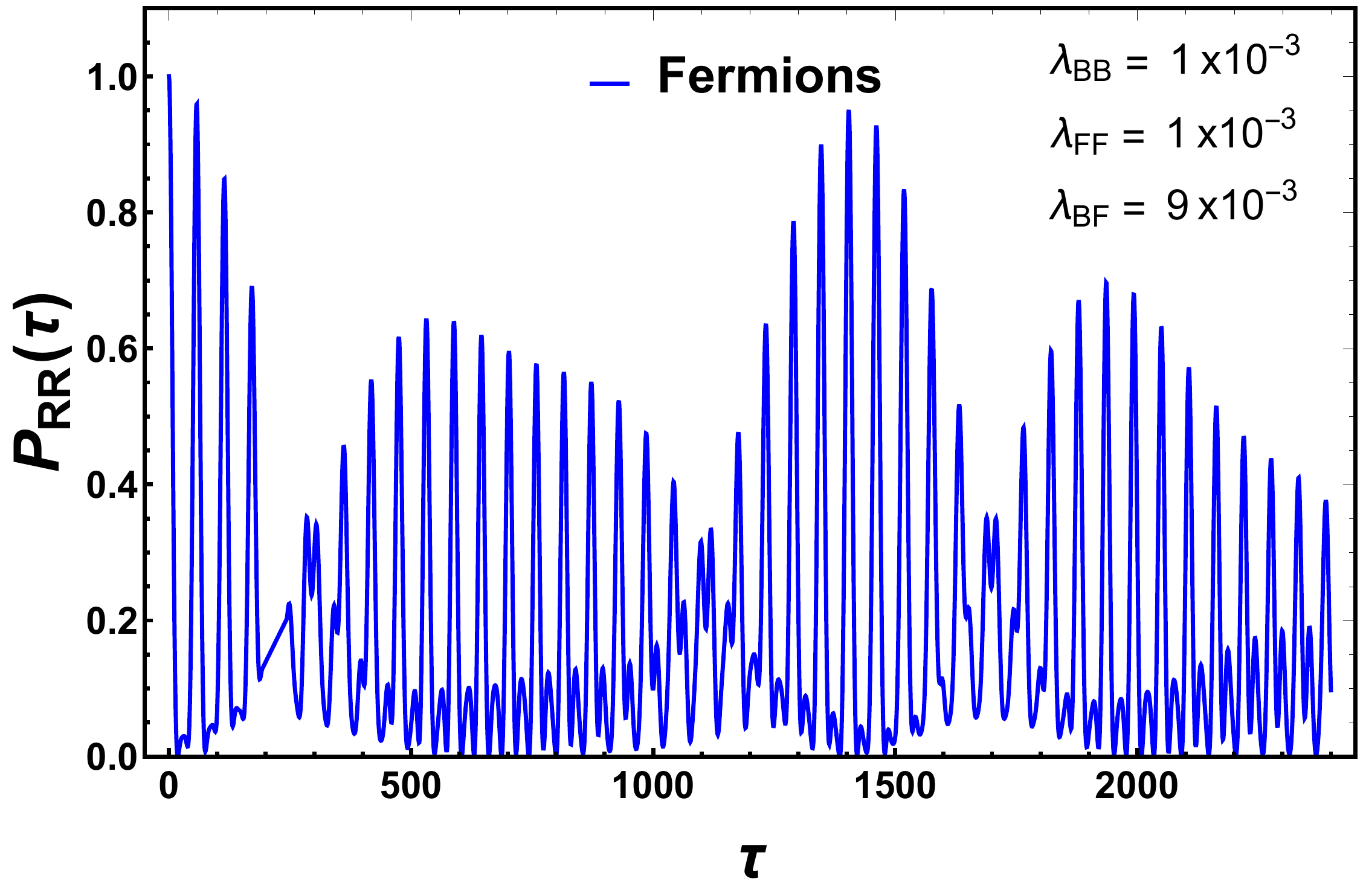}
\caption{Quasi periodicity for fermions. The graph shows $\tau$ for which the maximum occupancy per site is two fermions}
\label{fermionscuasi}
\end{subfigure}

\caption{Quasi-periodic probability density for (a) bosons,  red line and (b) fermions, blue line, considering $\lambda_{BB}=1\times10^{-3}, \lambda_{FF}=1\times10^{-3}$ and $\lambda_{BF}=9\times10^{-3}$.}
\label{cuasi}
\end{figure}

In the third region we find an increasing in the tunneling frequency of the two species, characterized by a higher damping of the probability densities and a higher return quasi periodicity for bosons (red line) than for fermions (blue line), as shown in figures \ref{bosonscuasi} and \ref{fermionscuasi} respectively. In the figure it can be seen that the probability density as a function of dimensionless time for the two species overlap in some regions, indicating a partial miscibility between bosons and fermions. It is also possible to find times for which the two bosons and the two fermions coexist in the same side of the potential well, as well as times for which sequential tunneling of bosons occurs.

By comparing the results of the figures \ref{constant2} and \ref{cuasi} it can be concluded that, the tunneling of bosons and fermions is sequential for small values of the interaction constant and correlated for large values.

\section{Quantum critical phenomena}\label{sec:QuaCriPhe}
The ground-state phase diagram of the model, is now introduced briefly through quantum fidelity and quantum entanglement concepts.

\begin{figure*}[htbp]
\begin{center}
  \subfloat[Fidelity in the $\lambda_{FF}-\lambda_{BF}$ plane and $\lambda_{BB}=5\times10^{-4}$.]{
   \label{f:bcte}
    \includegraphics[width=0.5\textwidth]{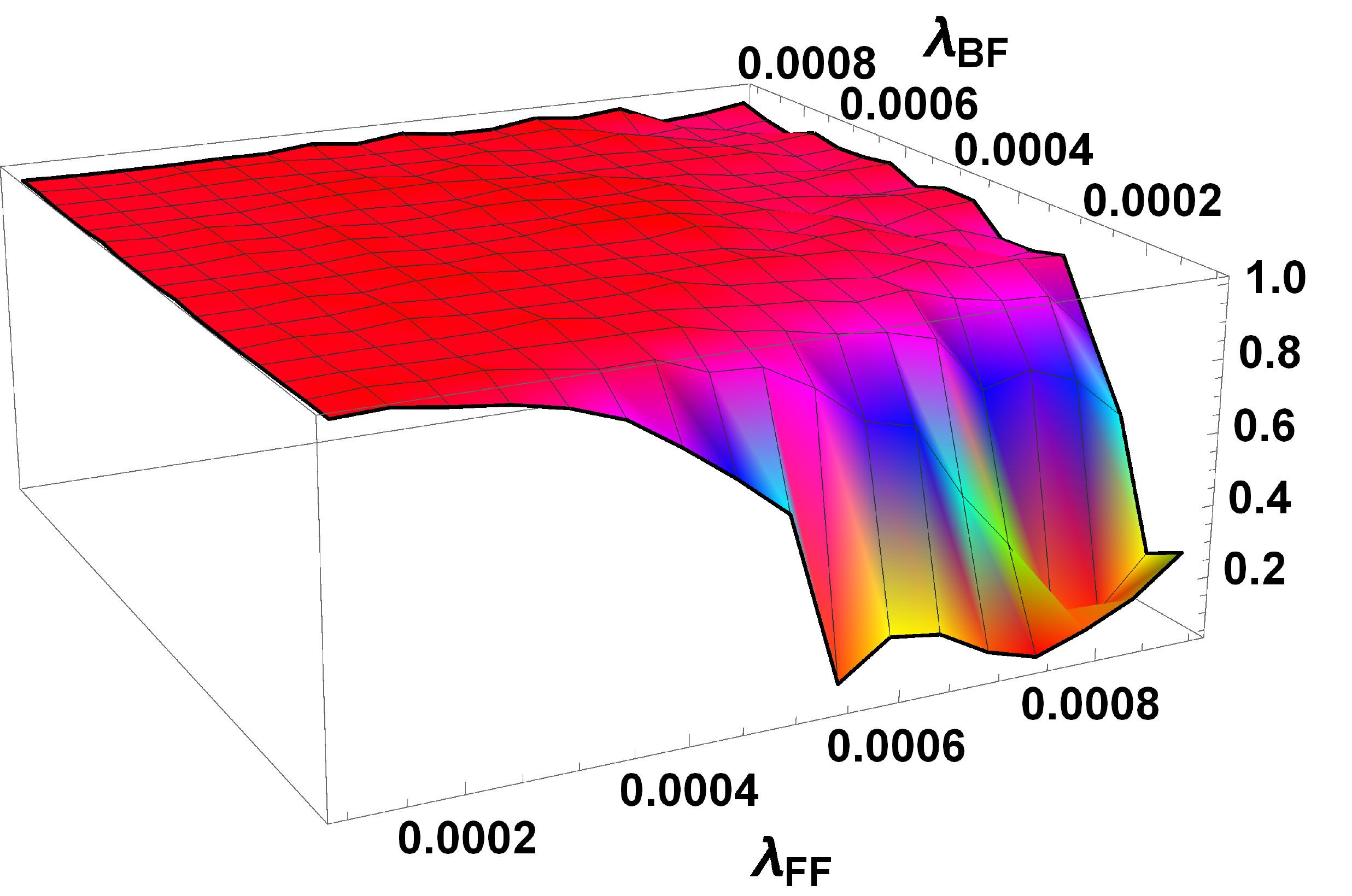}}
  \subfloat[Fidelity in the $\lambda_{BB}-\lambda_{BF}$ plane and $\lambda_{FF}=5\times10^{-4}$.]{
   \label{f:fcte}
    \includegraphics[width=0.5\textwidth]{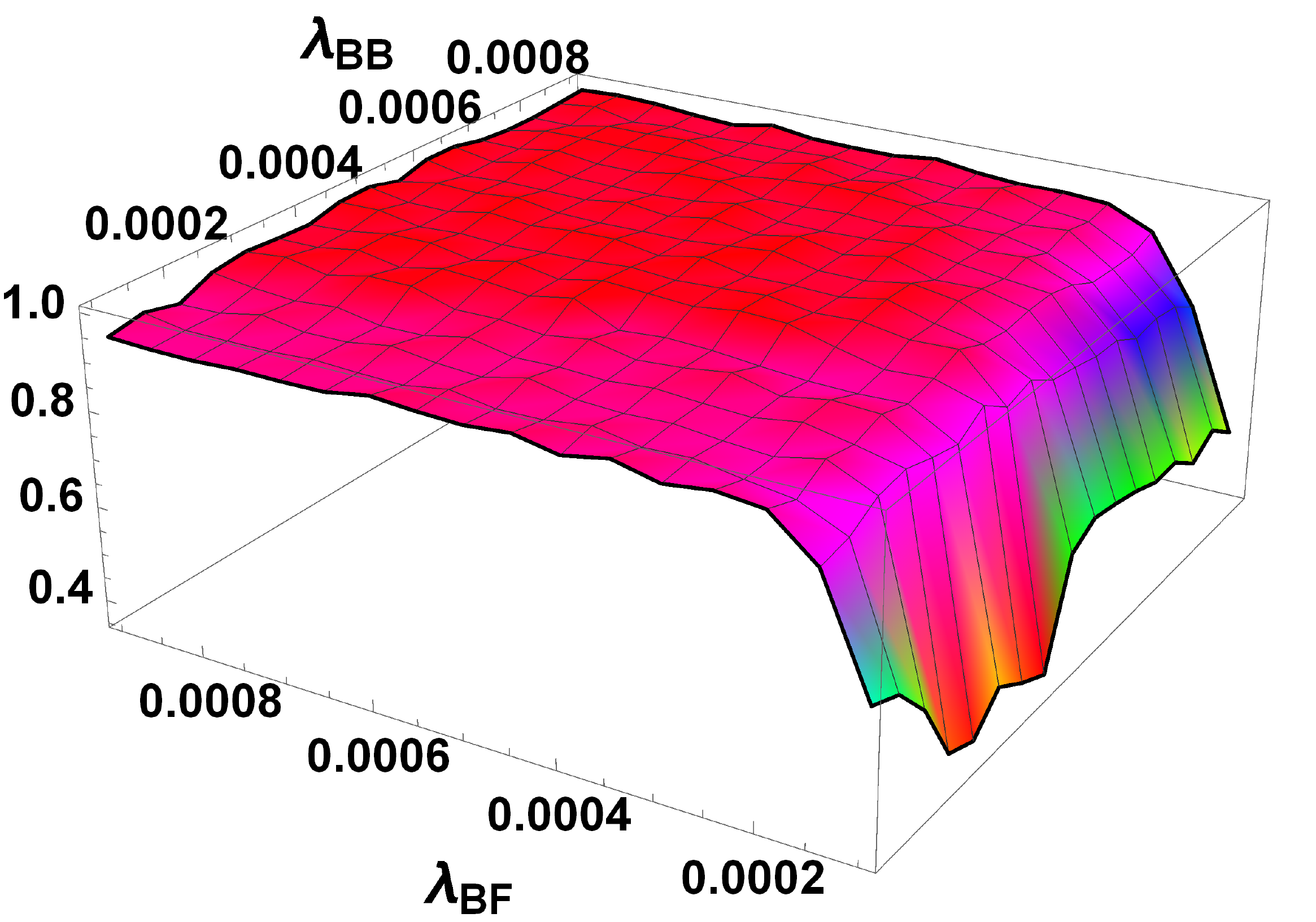}}\vspace{2mm}
  \subfloat[Fidelity in the $\lambda_{BB}-\lambda_{FF}$ plane and $\lambda_{BF}=5\times10^{-4}$.]{
   \label{f:bfcte}
    \includegraphics[width=0.5\textwidth]{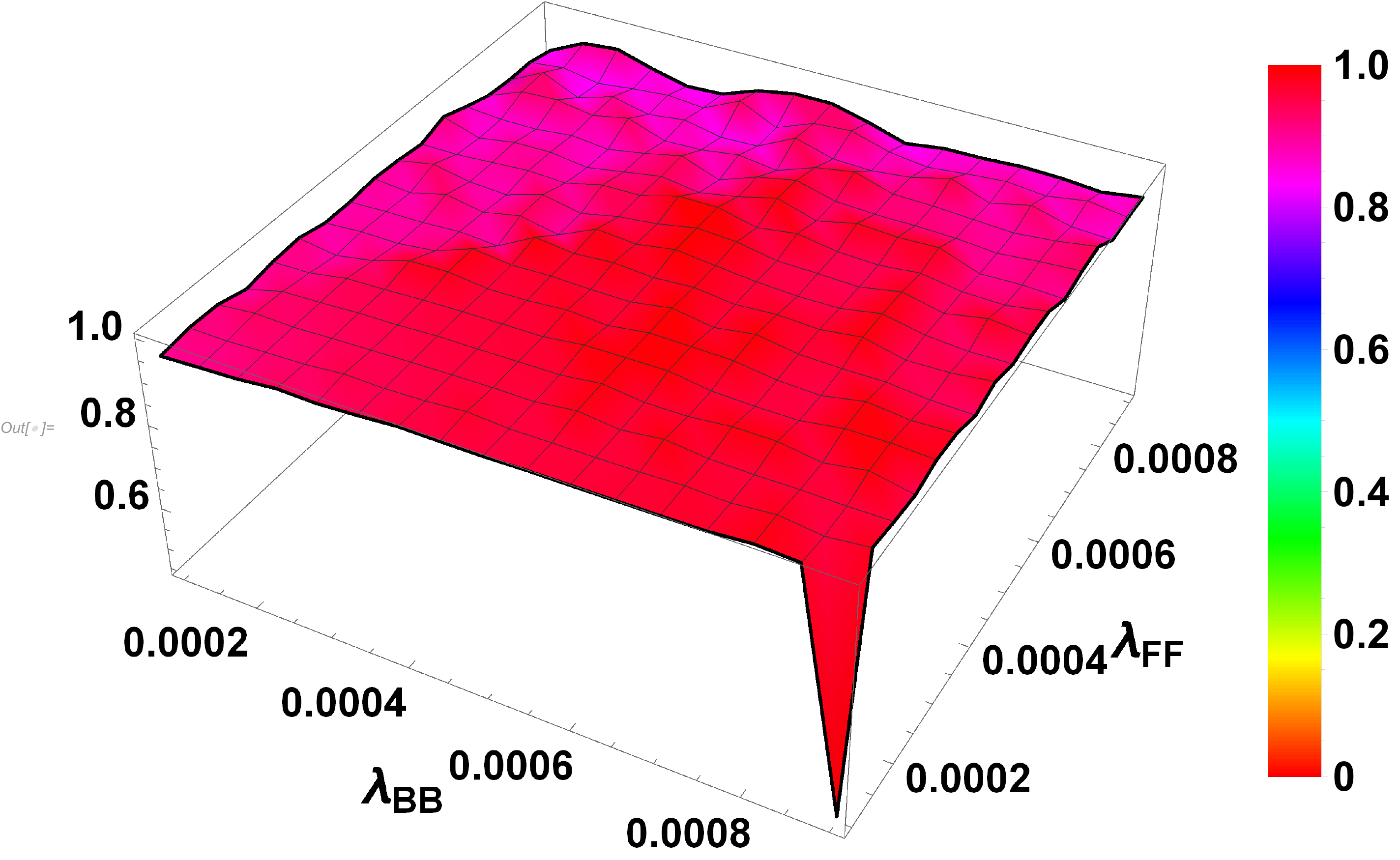}}
 \caption{The ground-state phase diagram of the model. It can be observed that in the three planes, there are two quantum phase transition indicated by the drop in the fidelity.}
 \label{f:fidelity}
\end{center}
\end{figure*}
Firts we to introduce the measure distance between two states trought quantum fidelity, that is very useful in the context of teleportation and quantum cryptography and is defined as
\begin{equation}
\label{fidelity}
F(|\psi\rangle,|\phi\rangle)=|\langle\psi|\phi\rangle|
\end{equation}
in wich $|\psi\rangle$ and $|\phi\rangle$ represent the ground states of the hamiltonian \ref{hubofear} with different parameters and it is nothing but the modulus of the overlap of two ground states generated by lightly different parameters and whose drop is a signature of a quantum phase transition (QPT).

In this research we set the ground state of the system at $\lambda_{BB}=\lambda_{FF}=\lambda_{BF}=5\times10^{-4}$ and calculate the fidelity in the $\lambda_{FF}-\lambda_{BF}$ plane, setting the value of the interaction between bosons at $\lambda_{BB}=5\times10^{-4}$, as shown in fig. \ref{f:bcte}, where for values of the interaction parameters $0\leq\lambda_{FF}\leq10\times10^{-4}$ and $2\times10^{-4}\leq\lambda_{BF}\leq1\times10^{-3}$  the fidelity is approximately constant around $1$, and there is a drop in fidelity when the values of the interactions are between $3.5\times10^{-4}\leq\lambda_{FF}\leq1\times10^{-3}$ and $0\leq\lambda_{BF}\leq2\times10^{-4}$

In order to know the fidelity in the $\lambda_{BB}-\lambda_{BF}$ and $\lambda_{BB}-\lambda_{FF}$ planes, the interaction parameters $\lambda_{FF}=5\times10^{-4}$ and $\lambda_{BF}=5\times10^{-4}$ were set as shown in figures  \ref{f:fcte} and \ref{f:bfcte} respectively. We would like to emphasis that these phase diagrams are obtained without any knowledge of the correlation properties of the system.

On figures \ref{f:bcte} and \ref{f:fcte} is also clearly observed that the drop of the two fidelities becomes deeper and deeper as the boson-fermion interaction decreases. This
phenomenon indicates that although the phase transition is within the same class but the similarity of the ground state is changing, as a consequence of the variation of the parameter $\lambda_{BF}$ which induces a repulsive interaction between fermions, due to the interspecies entanglement. A similar result was reported for a mixture of spinless bosons and spinless fermions confined in a one-dimensional optical lattice \cite{PhysRevLett.96.190402, Ning_2008}.

On the other hand, in the $\lambda_{BB}-\lambda_{FF}$ plane it is observed that the drop in fidelity occurs for values of the interaction parameter between fermions very close to zero, as shown in figure\ref{f:bfcte}.

We have discussed quantum fidelity, now  it would be apt to introduce the  entanglement.  Because our system is configured by identical particles, it is necessary to take into account that the correlations arising are due to purely from their indistinguishable nature and should be excluded from the definition of quantum correlations, that are closely related to quantum entanglement \cite{ECKERT200288, PhysRevA.70.012109}. For the above reason to measure the amount of entanglement, the symmetrized (bosons) and anti-symmetrized (fermions) products of one-particle functions are considered as non-entangled and deviations from such states are used to measuring the amount of correlation \cite{Okopi_ska_2010}.

There exist different quantitative measures of amount of entanglement, for example, Schimdt measure, entanglement of formation, concurrence, negativity, etc. Here we measure the amount of entanglement through the Von Neumann entropy, which is defined as

\begin{equation}
\label{VonNeu}
S(\rho)=-Tr(\rho log_2 \rho)=-\sum_i \lambda_ilog_2\lambda_i,
\end{equation}

where $\rho$ represent the density matrix and $\lambda_i$ are its eigenvalues.

\begin{figure}[H]
\centering
\includegraphics[width=\linewidth]{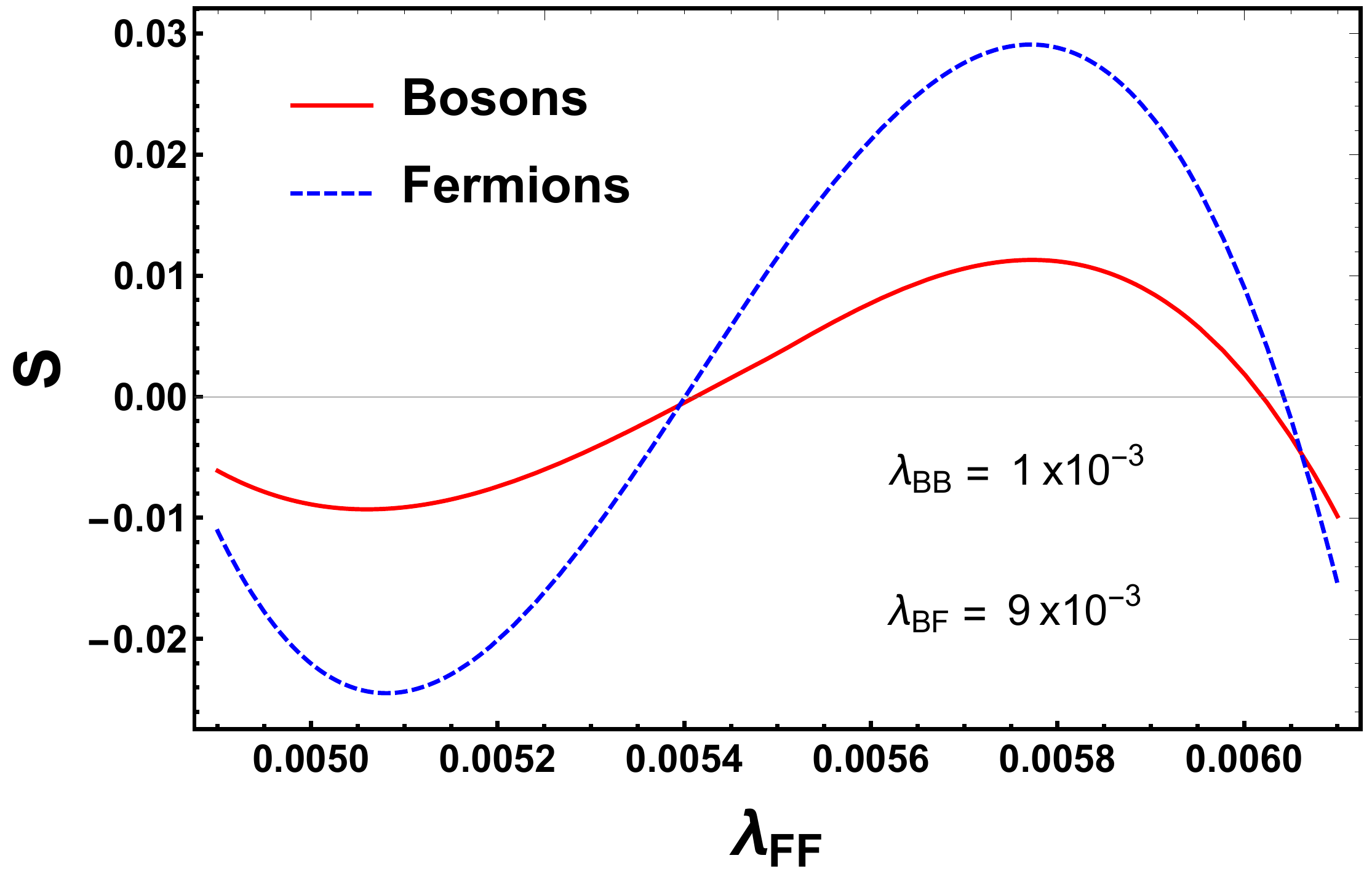}
\captionsetup{justification=justified}
\caption{Von Neumann as a function of $\lambda_{FF}$. Quantum correlations of bosons (red line) and fermions (blue line). The fastest entanglement growth takes place at $\lambda_{FF}=5.7\times10^{-3}$ for both species.}
\label{BosFerQuaentr}
\end{figure}

Due to the great sensitivity of the system to variations of the interaction parameters, we set the boson-boson interaction and the boson-fermion interaction at $1\times10^{-3}$ and $9\times10^{-3}$ respectivily and we explore a small region in which the boson-fermion coupling allows a maximum entanglement of the system as shown in figure  \ref{BosFerQuaentr}, where  quantum correlations of bosons (red line) and fermions (blue line) are presented by means of the calculation of the Von Neumann entropy $S(\rho)$ \ref{VonNeu} as a function of fermion-fermion interaction. The fastest entanglement growth takes place at $\lambda_{FF}=5.7\times10^{-3}$  where entropy is maximal.

\section{SUMMARY AND CONCLUSION}\label{sec:Conclusions}

We have investigated the time evolution of probability density, the ground-state fidelity and the entanglement of a Bose-Fermi mixture in a 1D double well potential, considering spinless bosons in the soft-core limit and spin one half fermions with inter and intra particles weak repulsive contact interaction.

Tuning the ratio between the inter and intraspecies interaction strengths, we found that the behaviour of the quantum return probability $P_{RR}$ present three distinct regions. The first region is characterized by correlated tunneling of bosons and fermions, where probabilities are periodic and corresponds to very small values of the effective coupling constants. In this region the bosonic and fermionic probability densities exhibit complete miscibility, characterized by an overlapping of the probability densities of the two species as a function of time. The second region indicates that particles of both species tend to separate and occupy both sides of the confinement potential and gives rise to correlated sequential tunneling, characterized by one particle of each species on each side of the well. In the last region we find an increasing in the tunneling frequency of the two species, characterized by a higher damping of the probability densities and a higher return quasi periodicity for bosons than for fermions. The above results allow us to conclude that, the tunneling of bosons and fermions is sequential for small values of the interaction constant and correlated for large values.

The ground-state phase diagram of the model, was briefly studied through of quantum entanglement and quantum fidelity concepts. We found through the Von Neumann entropy S as a function of fermion-fermion interaction, that the boson-fermion coupling allows a maximum entanglement of quantum correlations of bosons and fermions in the same value.

Finally we calculated the fidelity in the $\lambda_{FF}-\lambda_{BF}$ and $\lambda_{BB}-\lambda_{BF}$  planes and we found that the
drop of the two fidelities becomes deeper and deeper as the boson-fermion interaction decreases. This phenomenon indicates that although the phase transition is within the same class but the similarity of the ground state is changing, as a consequence of the variation of the parameter $\lambda_{BF}$.

Our findings can help to interpret experimental results in bosonic Josephson junction, squeezing, entanglement of matter waves, matter wave interference and in quantum information processing.

\section{ACKNOWLEDGMENTS}

R. A. and J.P.R. are thankful for the support of Departamento Administrativo de Ciencia, Tecnología e Innovación (COLCIENCIAS) (Convocatoria Doctorados Nacionales 727 of 2015 and 567 of 2012 respectively.) and  gratefully acknowledge support from "Fundación Universitaria los libertadores".

D. G. gratefully acknowledge support from " Universidad EAN".

\bibliography{apssamp}

\end{document}